\documentstyle[12pt]{article}
\newcommand{\EQ}{\begin{equation}}
\newcommand{\EN}{\end{equation}}
\newcommand{\bea}{\begin{eqnarray}}
\newcommand{\eea}{\end{eqnarray}}

\newcommand{\th}{\theta}

\newcommand{\goto}{\rightarrow}
\newcommand{\lab}{\label}

\newcommand{\zb}{\bar{z}}

\begin{document}
\setcounter{page}{0}
\topmargin 0pt
\oddsidemargin 5mm
\renewcommand{\thefootnote}{\arabic{footnote}}
\newpage
\setcounter{page}{0}
\begin{titlepage}
\begin{flushright}
OUTP-96-25S\\
SWAT/95-96/123
\end{flushright}
\vspace{0.5cm}
\begin{center}
{\large {\bf Asymptotic factorisation of form factors in
two-dimensional quantum field theory}}
\end{center}
\vspace{1.8cm}
\begin{center}
{\large G. Delfino$^{1}$, P. Simonetti$^{2}$ and J.L. Cardy$^{1}$}
\end{center}
\vspace{0.5cm}
\begin{center}
{\em $^{1}$ Theoretical Physics, University of Oxford\\
            1 Keble Road, Oxford OX1 3NP, United Kingdom}\\
{\em $^{2}$ Department of Physics, University of Wales Swansea\\
            Singleton Park, Swansea SA2 8PP, United Kingdom}
\end{center}   
\vspace{0.5cm}
\begin{center}
emails: g.delfino1@physics.oxford.ac.uk, j.cardy1@physics.oxford.ac.uk \\ 
p.simonetti@swansea.ac.uk
\end{center}   

\vspace{1.2cm}

\renewcommand{\thefootnote}{\arabic{footnote}}
\setcounter{footnote}{0}

\begin{abstract}
\noindent
It is shown that the scaling operators in the conformal limit of a
two-dimensional field theory have massive form factors which obey a
simple factorisation property in rapidity space. This has been used to
identify such operators within the form factor bootstrap approach. A
sum rule which yields the scaling dimension of such operators is also
derived. 
\end{abstract}

\vspace{.3cm}

\end{titlepage}

\newpage
\noindent
{\bf 1.} The solution of conformal field theories represented a crucial step
in our understanding of two-dimensional quantum field theories
\cite{BPZ,saleur}. 
It not only amounts to a complete description of the fixed
points of the renormalisation group (RG), but also provides the
starting point for the study of physical systems away from
criticality. For example, one can
consider the theory defined by the action
\EQ
{\cal A}={\cal A}_{CFT}+g\int d^2x\,\varphi(x)\,\,,
\lab{action}
\EN
as describing the perturbation of a conformal invariant theory by the
operator $\varphi(x)$ of scaling dimension $2\Delta<2$. 
Since the coupling constant $g$ has physical dimension
$m^{2-2\Delta}$, $m$
being a mass scale, the theory (\ref{action}) is no longer scale
invariant; rather, it is associated to a RG trajectory flowing out
of the original fixed point.
In many cases an infinite number of integrals of motion  survive in the
perturbed theory and the resulting off-critical model is
said to be integrable. A bootstrap procedure can then be applied,
usually resulting in the determination of the exact particle spectrum
and $S$-matrix of the theory \cite{Zam&Zam,taniguchi,report}. As a consequence
of integrability, the $S$-matrix turns out to be completely elastic
and factorised. On the other hand, it is commonly believed that the
knowledge of the $S$-matrix amounts to a complete solution of a
quantum field theory. In particular, it should encode the information
about the operator content of the model and should enable the
computation of correlation functions. The method which has proved so far
very effective in dealing with such an ambitious program is known
as the {\em form factor bootstrap}. Form factors (FF) are matrix elements
of local operators ${\cal O}(x)$ between asymptotic multiparticle
states and will be denoted as \footnote{In order to avoid inessential
complication of the notation, we refer throughout this letter to a
theory whose spectrum consists of a single species $A$ of particles
with mass $m$. We also adopt the standard parameterisation of the on
mass shell momenta in terms of the rapidity variables $\th_i$:
$p_i^\mu=(m\cosh\th_i,m\sinh\th_i)$.} 
\EQ
F_n^{\cal O}(\th_1,\ldots,\th_n)=\langle
0|{\cal O}(0)|A(\th_1),\ldots,A(\th_n)\rangle\,\,\,.
\lab{ff}
\EN
They are an interesting subject for theoretical investigation
because their structure involves the particle description of the
theory, an expression of the infrared dynamics, as well as the operator
content, which is deeply related to the conformal structure of the
ultraviolet fixed point. Moreover, if the FF are known, correlation
functions can be written down in terms of the spectral sum
\EQ
\langle{\cal O}_1(x){\cal
O}_2(0)\rangle=\sum_n\frac{1}{n!}\int\frac{d\th_1}{2\pi}\ldots
\frac{d\th_n}{2\pi}F_n^{{\cal O}_1}(\th_1,\ldots,\th_n)
\left[F_n^{{\cal O}_2}(\th_1,\ldots,\th_n)\right]^*
e^{iP^\mu_nx_\mu}\,\,,
\lab{spectral}
\EN
$P_n^\mu$ being the total energy-momentum of the $n$-particle
intermediate state.

The determination of FF in integrable models proceeds through two
basic steps. One first enforces the general constraints of
analyticity, unitarity and crossing symmetry deriving from the
standard $S$-matrix theory \cite{karowski,smirnov}. Since no
information about the specific nature of the operator ${\cal O}(x)$ is
provided at this stage, the general solution of the resulting system
of functional and residue equations (linear in the operator) must
correspond to a complete description of the operator content of the
theory. It has been shown for several models that the dimensionality
of the linear space of solutions of these equations coincides with
that predicted by the underlying conformal theory
\cite{CM1,annicounting,sgcounting}. 

The second step consists in selecting out of the general solution the
FF corresponding to particular operators, the scaling operators being the
objects of main physical interest. Such operators provide the
natural physical basis for the space of local operators of the
theory. Nevertheless, it is by no means obvious how to identify this
basis among the solutions of the FF equations. A first selection
rule comes from the general constraint \cite{immf}
\EQ
\lim_{|\th_i|\goto \infty}F_n^\Phi(\th_1,\ldots,\th_n)\leq
const. \,\,e^{\Delta_\Phi|\th_i|}\,\,,
\lab{bound}
\EN
relating the asymptotic behaviour of the matrix elements of a scaling
operator $\Phi(x)$ to its scaling dimension $2\Delta_\Phi$ \footnote{We will
refer to scalar operators.}. When the operator space splits into 
different sectors distinguished by some internal symmetry of the
theory, the asymptotic bound (\ref{bound}) is typically sufficient to
fix the FF solutions. Yet, the problem remains conceptually interesting
in absence of internal symmetries. We now show that in the latter case
the FF of the relevant ($\Delta_\Phi<1$) scaling operators of the theory
are characterised by the following asymptotic factorisation property
\EQ
\lim_{\alpha\goto+\infty}F_{r+l}^\Phi(\th_1+\alpha,\ldots,\th_r+\alpha,
\th_{r+1},\ldots,\th_{r+l})=\frac{1}{\langle\Phi\rangle}
F_r^\Phi(\th_1,\ldots,\th_r)\,F_l^\Phi(\th_{r+1},\ldots,\th_{r+l})\,\,\,.
\lab{cluster}
\EN
Such a property had been noticed in the past to be fulfilled by some FF
solutions in specific models \cite{cluster}.

Consider the case in which the action (\ref{action}) describes a massive
integrable model without internal symmetries and denote by $S(\th)$
the two-particle scattering amplitude, $\th$ being the rapidity
difference between the colliding particles. It
can be argued on general grounds that
\EQ
\lim_{\th\goto\infty}S(\th)=1\,\,,
\lab{slim}
\EN
and that this result implies 
that the FF of a relevant scaling operator $\Phi(x)$ actually tend 
to a constant
(with respect to $\th_i$) in the limit (\ref{bound}), so that the 
limit in eq.\,(\ref{cluster}) is surely well defined. Also,
due to the absence of internal symmetries, the vev
$\langle\Phi\rangle$ should be nonvanishing and can be written as
\EQ
\langle\Phi\rangle=v_\Phi\, m^{2\Delta_\Phi}\,\,,
\lab{vev}
\EN
$v_\Phi$ being a dimensionless constant.

The basic point to be realised is that the limit in the r.h.s. of 
eq.\,(\ref{cluster}) is actually a massless limit into the
ultraviolet conformal point. In general, such limit is obtained
sending to zero the mass of each particle while sending to infinity
its rapidity in order to keep the energy finite \cite{zamotim,zamoSU2}. 
The resulting theory
consists of right and left-moving massless particles, $A_R$ and $A_L$,
whose Zamolodchikov-Faddeev operators may be 
formally obtained from those of the original massive particle $A$ as
\bea
& A_R(\theta)=\lim_{\alpha\goto+\infty}A(\theta+\alpha/2)\,\,,\\
\nonumber
& A_L(\theta)=\lim_{\alpha\goto+\infty}A(\theta-\alpha/2)\,\,.
\lab{massless}
\eea
The dispersion relations are
\EQ
p^0=p^1=\frac{M}{2}e^\theta\hspace{1cm}\mbox{for right-movers}\,\,,
\nonumber
\EN
\EQ
p^0=-p^1=\frac{M}{2}e^{-\theta}\hspace{1cm}\mbox{for left-movers}\,\,,
\EN
$M=me^{\alpha/2}$ being a finite parameter. The scattering amplitudes
characterising the interaction of the masseless particles are readily
obtained taking the limit of the massive amplitude:
$S_{RR}(\theta)=S_{LL}(\th)=S(\theta)$ and
$S_{RL}(\theta)=\lim_{\alpha\goto+\infty} S(\theta+\alpha)$. 
In the present context the latter limit for $S_{RL}$ 
gives a constant phase whose rapidity independence
ensures the decoupling of the right and left sectors and 
the scale invariance of the massless theory \footnote{When the
right-left amplitude is rapidity dependent the theory describes a
massless flow between two fixed points.}. In absence of internal
symmetries, eq.\,(\ref{slim}) implies $S_{RL}=1$.
In summary, the massless
scattering theory defined through the above limiting procedure provides a
particle description of the conformal point to which the original
massive model flows in the ultraviolet limit.

Massless form factors can be introduced along the same lines
\cite{DMSmassless}. In order to prevent a trivial vanishing of the
matrix elements (\ref{ff}) when $m\goto 0$, it is convenient to refer
to the rescaled operator
\EQ
\hat\Phi=\frac{\Phi}{m^{2\Delta_\Phi}}\,\,.
\EN
Since the matrix elements of a scalar operator depend on rapidity
differences only, we can write
\bea
& \lim_{\alpha\goto+\infty}F_{r+l}^{\hat\Phi}(\th_1+\alpha,\ldots,\th_r+\alpha,
\th_{r+1},\ldots,\th_{r+l})= \nonumber \\
& \lim_{\alpha\goto+\infty}F_{r+l}^{\hat\Phi}(\th_1+\alpha/2,\ldots,
\th_r+\alpha/2,\th_{r+1}-\alpha/2,\ldots,\th_{r+l}-\alpha/2)= \nonumber\\
& F_{r,l}^{\hat\Phi}(\th_1,\ldots,\th_r|\th_{r+1},\ldots,\th_{r+l})\,\,,
\lab{limit}
\eea
where
\EQ
F_{r,l}^{\hat\Phi}(\th_1,\ldots,\th_r|\th_{1}',\ldots,\th_{l}')\equiv
\langle
0|\hat\Phi(0)|A_R(\th_1),\ldots,A_R(\th_r),A_L(\th_{1}'),\ldots,
A_L(\th_{l}')\rangle\,\,,
\lab{masslessff}
\EN
denotes a massless form factor. On the other hand, as a consequence of
the decoupling of the right and left sectors at the conformal point,
the massless form factor  of a scaling operator can be written 
in the factorised form
\EQ
F_{r,l}^{\hat\Phi}(\th_1,\ldots,\th_r|\th_{1}',\ldots,\th_{l}')=
{\cal R}^{\hat\Phi}_r(\th_1,\ldots,\th_r)\,
{\cal L}^{\hat\Phi}_l(\th_{1}',\ldots,\th_{l}')\,\,.
\lab{factorisation}
\EN
Consider now a form factor in the massive theory and take its massless
limit in which all the particles become right-movers. Due to Lorentz
invariance, the rapidity shifts needed for the limit can be completely
rescaled out and we have
\EQ
F_{r}^{\hat\Phi}(\th_1,\ldots,\th_r)=
F_{r,0}^{\hat\Phi}(\th_1,\ldots,\th_r|)=
{\cal R}^{\hat\Phi}_r(\th_1,\ldots,\th_r)\,
{\cal L}^{\hat\Phi}_0\,\,.
\lab{ronly}
\EN
A similar equation holds when all the particles become
left-movers. Since $\langle\hat\Phi\rangle=v_\Phi$ along the whole
flow, at the critical point we have
\footnote{Remember that $\hat\Phi$ is a rescaled (dimensionless)
operator; the vev (\ref{vev}) of the operator $\Phi$ vanishes in the
massless limit, as expected in conformal field theory.}
\EQ
{\cal R}^{\hat\Phi}_0={\cal L}^{\hat\Phi}_0=v_\Phi^{1/2}\,\,.
\lab{rvev}
\EN
Using eqs.\,(\ref{factorisation}), (\ref{ronly}) and (\ref{rvev}),
eq.\,(\ref{limit}) can be rewritten as
\EQ
\lim_{\alpha\goto+\infty}F_{r+l}^{\hat\Phi}(\th_1+\alpha,\ldots,\th_r+\alpha,
\th_{r+1},\ldots,\th_{r+l})=\frac{1}{v_\Phi}
F_r^{\hat\Phi}(\th_1,\ldots,\th_r)\,
F_l^{\hat\Phi}(\th_{r+1},\ldots,\th_{r+l})\,\,,
\EN
in terms of massive FF only. Going back from $\hat\Phi$ to
$\Phi$, eq.\,(\ref{cluster}) follows.

It is worth noticing that eq.\,(\ref{cluster}) can be used to
determine the vev $\langle\Phi\rangle$ from the knowledge of the
multiparticle FF \footnote{More precisely, what one can fix is the
ratio $\frac{\langle\Phi\rangle}{F_1^\Phi}$ which, being independent
from the normalisation of the operator, is an universal number.}.
Taking into account that only the vev of the
perturbing operator can be obtained by other means (thermodynamic Bethe
ansatz), this must be regarded as a remarkable circumstance. 

The crucial conditions entering the above derivation of the
factorisation property (\ref{cluster}) are the nonvanishing of all 
the FF of the considered scaling operator (including the vev) and their 
constant asymptotic behaviour. These features, which are guaranteed
in absence of internal symmetries, are often shared by some operators
in theories with symmetry. The FF of these operators then factorise
asymptotically according to eq.\,(\ref{cluster}). As an example, we
mention the exponential operators in lagrangian theories like the
Sinh-Gordon model \cite{cluster}, or the linear combinations 
$\sigma\pm\mu$ of the order and disorder parameters in the thermal 
Ising model \cite{berg,zamoising}. It seems reasonable to expect that
a suitable generalisation of eq.\,(\ref{cluster}) should apply to
any scaling operator even in presence of internal symmetries, but no
general pattern has been identified so far.

\vspace{1cm}
\noindent
{\bf 2.} Once the FF solutions for the relevant scaling operators have been
selected using the factorisation property (\ref{cluster}), it remains
to be established which scaling operator each solution corresponds
to. For this purpose, it is important to be able to recover the
scaling dimension $2\Delta_\Phi$ from the FF solution. In principle
this can be extracted evaluating the short distance behaviour of the
spectral series (\ref{spectral}) for the correlator
$\langle\Phi(x)\Phi(0)\rangle$. In practice, better quantitative
results can be obtained exploiting the properties of the stress tensor
as the generator of dilatations. In order to illustrate this point it is
useful to start with some perturbative consideration. 

The operator space of the conformal point and that of the perturbed
theory have the same basic structure. In particular, a scaling 
operator $\Phi(x)$ in
the off-critical theory can be associated to a conformal operator 
$\tilde\Phi(x)$ of scaling dimension $2\Delta_{\Phi}$. When doing that
in a perturbative framework, however, renormalisation effects induced 
by ultraviolet divergences must be taken into account \cite{zamo-yl}. 
In fact, denote by $X$ a generic product of operators and consider the
usual perturbative expansion of the correlator 
\EQ
\langle X\,\Phi(0)\rangle=\langle X\,\tilde\Phi(0)\rangle_{CFT}
+g\,\int_{\epsilon<|x|<R}d^2x\,\langle X\,\tilde\Phi(0)\,\tilde\varphi(x)
\rangle_{CFT}+{\cal O}(g^2)\,\,,
\lab{perturb}
\EN
where the correlators in the right hand side are computed in the
conformal theory, and $\epsilon$ and $R$ regularise the ultraviolet and
infrared divergences, respectively. The integral in
eq.\,(\ref{perturb}) is UV divergent only if the conformal OPE
\EQ
\tilde\varphi(x)\,\tilde\Phi(0)=\sum_k
C_{\varphi\Phi}^k\,|x|^{2(\Delta_k-\Delta_\Phi-\Delta)} \,\tilde A_k(0)
\lab{ope}
\EN
contains operators $\tilde A_k$ with scaling dimension $2\Delta_k$ such
that
\EQ
\gamma_k\equiv\Delta_k-\Delta_\Phi-\Delta+1\leq 0\,\,\,.
\lab{gamma}
\EN
In this case we obtain a first order UV finite correlator $\langle
X\,\Phi(0)\rangle$ by defining the renormalised operator as 
\footnote{Eq.\,(\ref{renorm}) must be suitably modified in the case 
of logarithmic divergences ($\gamma_k=0$).}
\EQ
\Phi=\tilde\Phi+g\sum_k b_k\epsilon^{2\gamma_k}\,\tilde A_k
+{\cal O}(g^2)\,\,,
\lab{renorm}
\EN
where
\EQ 
b_k=-\frac{\pi C_{\varphi\Phi}^k}{\gamma_k}\,\,,
\EN
and the sum runs only over the operators $\tilde A_k$ satisfying the
condition (\ref{gamma}). This implies that, in general,
renormalisation mixes the original operator $\tilde\Phi$ with a finite
number of operators of less scaling dimension.

Consider now the euclidean correlators of the operator $\Phi$ with
the components of the stress energy tensor
$T=\frac{1}{4}(T_{11}-T_{22}-2iT_{12})$ and $\Theta=T_{11}+T_{22}$
\bea
&& \langle T(z,\zb)\Phi(0)\rangle=\frac{F(z\zb)}{z^2}\,\,,\nonumber \\
&& \langle \Theta(z,\zb)\Phi(0)\rangle_c=\frac{G(z\zb)}{z\zb}\,\,,
\eea
where $z$ and $\zb$ are complex coordinates. Conservation of the stress
energy tensor espressed by the equation
\EQ
\bar\partial T+\frac{1}{4}\partial\Theta=0\,\,,
\EN
leads to the differential relation
\EQ
\dot D=\frac{1}{4}G\,\,,
\lab{diff}
\EN
where $D\equiv F+\frac{1}{4}G$ and the dot stays for
$z\zb\frac{d}{dz\zb}$. Since the trace of the stress tensor is related
to the perturbing field as 
\EQ
\Theta(x)=4\pi g\,(1-\Delta)\,\varphi(x)\,\,,
\lab{trace}
\EN
the short distance behaviour of the
function $G$ is determined by the OPE (\ref{ope})
\EQ
G(x)\simeq 2\pi g(2-2\Delta)C_{\varphi\Phi}^0\,|x|^{2\gamma_0}\,
\langle A_0\rangle\,\,,\hspace{.8cm}x\goto 0
\lab{G}
\EN
where we denoted by $A_0$ the most relevant operator appearing in
(\ref{ope}). We can now distinguish two basic cases:

{\em a)} $\gamma_0>0$. In this case $G$ vanishes as $x\goto 0$ (conformal
limit) and we conclude that the function $D$ is stationary and
coincides with $F$ at the fixed point. Since
the operator $\Phi(x)$ does not mix under renormalisation, its conformal
OPE with $T(x)$ can be safely used in a neighbourhood of the fixed
point
\EQ
F(x)\simeq \Delta_\Phi\langle\Phi\rangle\,\,,\hspace{.8cm}x\goto 0\,\,.
\lab{F}
\EN
If the theory described by the action (\ref{action}) corresponds to a
massless flow between two fixed points, a similar analysis can be
repeated in the neighbourhood of the infrared fixed point. Then,
integrating eq.\,(\ref{diff}) over all distance scales, one finds
\EQ
\Delta_{\Phi}^{UV}-\Delta_{\Phi}^{IR}=-\frac{1}{4\pi\langle\Phi
\rangle}\int d^2x\,\langle\Theta(x)\,\Phi(0)\rangle_c\,\,.
\lab{sumrule}
\EN
In a massive theory $\Delta_\Phi^{IR}=0$. We will illustrate in a
moment with few examples the effectiveness of this sum rule within the
FF approach.

{\em b)} $\gamma_0\leq 0$. The function $G$ no longer vanishes at
$x=0$ and an attempt to use the sum rule (\ref{sumrule}) would be
frustrated by the divergence of the integral. This is a consequence of
the fact that the operator $\Phi(x)$ now mixes under renormalisation (see
eq.\,(\ref{renorm})) so that the function $F$ no longer behaves as in
(\ref{F}) at short distances. The correct behaviour can instead be
obtained integrating eq.\,(\ref{diff}). One finds
\EQ
F(x)\simeq\pi
g\,(1-\Delta)C_{\varphi\Phi}^0\frac{1-\gamma_0}{\gamma_0}\,|x|^{2\gamma_0}
\langle A_0\rangle\,\,,\hspace{.8cm}\gamma_0<0
\lab{Fpower}
\EN
\EQ
F(x)\simeq 2\pi g\,(1-\Delta)C_{\varphi\Phi}^0\log|x|
\langle A_0\rangle\,\,,\hspace{.8cm}\gamma_0=0\,\,.
\lab{Flog}
\EN

Actually, eq.\,(\ref{sumrule}) is nothing but the Ward identity expressing
the fact that $\Theta$ is the generator of scale transformations, and as
such a suitable generalisation is valid in arbitrary dimension. This may
be derived simply by modifying the arguments presented, for example,
in ref.~\cite{CardyLH}. Consider the effect of making a nonuniform 
infinitesimal RG transformation, which generates the coordinate change
$x^\mu\to {x'}^\mu=x^\mu+\delta x^\mu$, where $\delta x^\mu=\epsilon x^\mu$, 
corresponding to uniform scale transformation,
in the region between
two hyperspheres, of radii $R_1$ and $R_2$, while $\delta x^\mu=0$ in
the
regions $|x|<R_1'<R_1$ and $|x|>R_2'>R_2$. Within the thin shells 
$R_1'<|x|<R_1$ and $R_2<|x|<R_2'$, $\delta x^\mu$ is chosen to be an
arbitrary differentiable function which matches smoothly onto the other
regions. 
{}From the definition of the stress tensor, the change in the hamiltonian
(action) is 
\begin{equation}
\delta S=-(1/S_d)\int\partial^\mu\delta x^\nu T_{\mu\nu}d^d\!x,
\end{equation}
where $S_d$, the area of a $d$-dimensional sphere, is conventionally
included. This gives a contribution $\int_{R_1<|x|<R_2}\Theta\, d^d\!x$
from this region. Within the two shells, integrate by parts: the bulk
term vanishes by the conservation $\partial^\mu T_{\mu\nu}=0$ of the
stress tensor, leaving two surface integrals on $|x|=R_1$ and $R_2$
of the form $\int x^\mu T_{\mu\nu}dS^\nu$. Now consider
evaluating $\langle\Phi(0)\rangle$ with respect to this modified
hamiltonian. Since no change has been made for $|x|<R_1$, this quantity
is in fact invariant. Thus the first order changes from the bulk and
boundary terms should cancel. Expanding these out from the hamiltonian
to $O(\epsilon)$, the bulk term gives 
\begin{equation}
-{1\over S_d}\int_{R_1<|x|<R_2}\langle\Theta(x)\Phi(0)\rangle_c\,d^d\!x,
\end{equation}
while the boundary terms involve integrals over $\langle T_{rr}(x)\Phi(0)
\rangle$. In the limits $R_1\to0$ and $R_2\to\infty$ these may be
evaluated using the operator product expansion formulae \cite{Cardydnot2}
appropriate to the UV or the IR conformal theories respectively:
\begin{equation}\label{oped}
T_{\mu\nu}(x)\Phi(0)={dx_\Phi\over d-1}{x_\mu x_\nu-(1/d)x^2g_{\mu\nu}
\over |x|^{d+2}}\Phi(0)+\cdots,
\end{equation}
where $x_\Phi$ is the scaling dimension of $\Phi$. This then leads to
the result
\begin{equation}
(x_\Phi^{UV}-x_\Phi^{IR})\langle\Phi\rangle=
-{1\over S_d}\int\langle\Theta(x)\Phi(0)\rangle_c\, d^d\!x,
\end{equation}
which is the generalisation of (\ref{sumrule}), with $x_\Phi$ twice the
complex dimension $\Delta_\Phi$.

In the case when the right hand side diverges, it is not permissible to
take the limit $(R_1\to0,R_2\to\infty)$, and, as above, this may be seen
through the operator mixing which is exhibited in the operator product
expansion (\ref{oped}).

\vspace{1cm}
\noindent
{\bf 3.} The Ising model provides an interesting example for testing the ideas
discussed in this letter. The situation is particularly simple in the
purely thermal case, which in terms of the action (\ref{action})
corresponds to perturbing the Ising critical point by the energy density
operator $\varepsilon(x)$ of scaling dimension
$2\Delta_\varepsilon=1$ ($g\sim T-T_c$). The only other relevant
operator in the model is the local magnetisation $\sigma(x)$ of scaling
dimension $2\Delta_\sigma=1/8$. The theory can be described in terms
of a free majorana fermion of mass $m$, corresponding to a scattering
amplitude $S(\th)=-1$. The only nonvanishing FF of the components of
the stress tensor $T$ and $\Theta=2\pi g\varepsilon$ (bilinear in the
fermions) are
\bea
F_2^\Theta(\th_1,\th_2)&=&-2i\pi m^2\sinh\frac{\th_1-\th_2}{2}\,\,,
\nonumber\\
F_2^T(\th_1,\th_2)&=&\frac{i\pi}{2}m^2e^{\th_1+\th_2}
\sinh\frac{\th_1-\th_2}{2}\,\,.\nonumber
\eea
The correlators of $\sigma$ with $T$ and $\Theta$ vanish for symmetry
reasons at $T>T_c$. Alternatively, one can work at $T<T_c$, or exploit
duality and refer instead to the disorder operator $\mu$. Here we only
need its two-particle FF 
\EQ
F_2^\mu(\th_1,\th_2)=i\langle\mu\rangle\tanh\frac{\th_1-\th_2}{2}\,\,.
\nonumber
\EN
The following (euclidean) correlators are then obtained using
eq.\,(\ref{spectral}) 
\bea
\langle T(x)T(0)\rangle&=&\frac{m^4}{16}
\left(\frac{\bar{z}}{z}\right)^2 \left[K_1(m|x|) K_3(m|x|) - 
K_2^2(m|x|)\right]\,\,,\nonumber\\
\langle T(x)\Theta(0)\rangle&=&\frac{m^4}{4}\frac{\bar{z}}{z}
\left[K_1^2(m|x|)-K_0(m|x|)K_2(m|x|)\right]\,\,,\nonumber\\
\langle T(x)\mu(0)\rangle&=&\frac{\langle\mu\rangle}{16z^2}
e^{-2m|x|}\,\,,\nonumber\\
\langle\Theta(x)\Theta(0)\rangle_c &=& m^4\left[K_1^2(m|x|)-
K_0^2(m|x|)\right]\,\,,\nonumber\\
\langle\Theta(x)\mu(0)\rangle_c &=& -m^2\langle\mu\rangle
\left[\frac{e^{-2m|x|}}{2m|x|}+
{\mbox Ei}(-2m|x|)\right]\,\,,\nonumber
\eea
where $K_\nu$ are the modified Bessel functions and ${\mbox Ei}$ the 
exponential integral function 
${\mbox Ei}(-x)=-\int_x^{+\infty}dt \exp(-t)/t$. One can easily check
that the above correlators satisfy the differential equation
(\ref{diff}). Comparison with the short distance expectations
\bea
\langle T(x)T(0)\rangle &\simeq &\frac{c}{2z^4}\,\,,\nonumber\\
\langle T(x)\varepsilon(0)\rangle &\simeq &\frac{2\pi
g}{z^2}(1-\Delta_\varepsilon)\log|x|\,\,,\nonumber\\
\langle T(x)\mu(0)\rangle &\simeq &\frac{\Delta_\mu}{z^2}\langle\mu\rangle\,\,,
\hspace{5cm} |x|\goto 0\nonumber\\
\langle\varepsilon(x)\varepsilon(0)\rangle &\simeq &
|x|^{-4\Delta_\varepsilon}\,\,,\nonumber\\
\langle\varepsilon(x)\mu(0)\rangle &\simeq & C_{\varepsilon\mu}^\mu
\langle\mu\rangle |x|^{-2\Delta_\varepsilon}\,\,,\nonumber
\eea
gives $c=1/2$, $\Delta_\mu=1/16$,
$\Delta_\varepsilon=1/2$, $m=2\pi g$ and
$C_{\varepsilon\mu}^\mu=-1/2$. In writing the second equation above we
took into account that $\gamma_0=0$ for $\varepsilon$ and used
eq.\,(\ref{Flog}) with $A_0=I$. Of course $\Delta_\mu$ can also be
obtained inserting the exact correlator
$\langle\Theta(x)\mu(0)\rangle_c$ into the sum rule (\ref{sumrule}).

Less simple is the case of the purely magnetic perturbation of the
Ising critical point ($\varphi(x)=\sigma(x)$ in the action
(\ref{action})). The theory is again integrable but the spectrum now
consists of eight massive particles \cite{taniguchi}. Since the presence of
the magnetic field breaks the invariance under spin reversal, no
internal symmetries are left in the model. The FF bootstrap program 
described in the first part of this letter was carried out for this
model in
ref.\,\cite{immf,energy}. In particular, the factorisation equation
(\ref{cluster}) was used in ref.\,\cite{energy} to identify the FF
solutions for the two relevant scaling operators $\sigma$ and
$\varepsilon$. Their scaling dimensions were also effectively estimated
using the sum rule (\ref{sumrule}).
Applications of the factorisation property (\ref{cluster}) to other
models are presented in ref.\,\cite{AMV}.

Finally, as an example of the massless case, let us consider the flow
from the tricritical to the critical Ising points (the minimal models
${\cal M}_{4,5}$ and ${\cal M}_{3,4}$ of conformal field theory,
respectively). The model is integrable and can be described in terms
of a single species of massless particles whose scattering is
characterised by the amplitudes $S_{RR}=S_{LL}=-1$ and
$S_{RL}(\th)=\tanh\left(\frac{\th}{2}-\frac{i\pi}{4}\right)$
\cite{zamotim}. The correlator $\langle\Theta(x)\mu(0)\rangle_c$ can
be computed through the spectral series (\ref{spectral}) using the
results of ref.\,\cite{DMSmassless}, where 
FF of several operators of the theory were determined.
Here we again consider for symmetry reasons the disorder parameter
$\mu$ instead of the magnetisation $\sigma$. Both operators have scaling
dimensions $3/40$ in the ultraviolet limit and $1/8$ in the infrared
limit. Being the most relevant operator in the theory, $\mu$ does not mix
under renormalisation and the sum rule (\ref{sumrule}) can be safely used to
evaluate the total variation in its scaling dimension along the
flow. The integration of the first (four-particle) contribution gives
$-0.0255$; the addition of the second (six-particle) contribution 
leads to the result $-0.0249$, showing that the FF series rapidly
converges to the expected result
$\Delta_\mu^{UV}-\Delta_\mu^{IR}=3/80-1/16=-0.025$ in spite of the
massless nature of the theory.

\vspace{1cm}
{\em Acknowledgements.} J.C. and G.D. were supported by the EPSRC
grant GR/J78044; P.S. was supported by a HEFCW grant.

\end{document}